\documentclass[conference]{IEEEtran}
\IEEEoverridecommandlockouts
 
\usepackage{listings}
\usepackage{color}

\definecolor{lightgray}{gray}{0.9}

\usepackage{enumitem,kantlipsum} 
\usepackage{placeins}
\usepackage{makecell, multirow}
\usepackage{mwe}
\usepackage{subcaption}
\usepackage{multicol}
\usepackage{booktabs,tabularx}
\usepackage{amsmath,amssymb,amsfonts}
\usepackage{float}
\usepackage{biblatex}
\usepackage[utf8]{inputenc} 
\usepackage{fancyhdr}
\usepackage{blindtext}
\usepackage{csquotes}
\usepackage[misc]{ifsym}  
\usepackage{braket}
\usepackage[
  separate-uncertainty = true,
  multi-part-units = repeat
]{siunitx}

\usepackage[T1]{fontenc}     

\usepackage{url}
\usepackage{hyperref}
\usepackage{cleveref }
\usepackage{algorithmic}
\usepackage{algorithm}
\usepackage{graphicx}
\usepackage{textcomp}
\usepackage{xcolor}    
\usepackage{caption}
\usepackage{tikz}
\usepackage{pgfplots, pgfplotstable}
\pgfplotsset{compat=1.17}

\usepackage{tabularx,booktabs}
\newcolumntype{C}{>{\centering\arraybackslash}X} 
\usepackage{lipsum}
\def\BibTeX{{\rm B\kern-.05em{\sc i\kern-.025em b}\kern-.08em
    T\kern-.1667em\lower.7ex\hbox{E}\kern-.125emX}}
     
\begin{document}

\title{Emotion Analysis of Social Media Bangla Text and Its Impact on Identifying the Author's
Gender}

\author{
\IEEEauthorblockN{Sultan Ahmed}
\IEEEauthorblockA{\textit{Department of Information Systems} \\
\textit{University of Maryland Baltimore County}\\
Maryland, USA \\
IL66977@umbc.edu}
\and 
\IEEEauthorblockN{Salman Rakin}
\IEEEauthorblockA{\textit{Department of CSE} \\
\textit{Bangladesh University }\\
\textit{ of Engineering \& Technology}\\
Dhaka, Bangladesh \\
0417052033@grad.cse.buet.ac.bd}
\and
\IEEEauthorblockN{Khadija Urmi}
\IEEEauthorblockA{\textit{Department of CSE} \\
\textit{BRAC University}\\
Dhaka, Bangladesh \\
khadija.urmi.cse@gmail.com}
\and 

\IEEEauthorblockA{\hspace*{9cm} 
\IEEEauthorblockN{Chandan Kumar Nag}
\textit{Department of CSE} \\
\textit{Southeast University}\\
Dhaka, Bangladesh \\
cknag11@gmail.com}

\and 
\IEEEauthorblockN{Dr. Md. Mostofa Akbar}
\textit{Department of CSE} \\
\textit{Bangladesh University }\\
\textit{ of Engineering \& Technology}\\
Dhaka, Bangladesh \\
mostofa@cse.buet.ac.bd}

\maketitle

\def\sbng{\bngviii}
\def\tbng{\bngvi}
\def\bng{\bngx}
\def\lbng{\bngxiv}
\def\Lbng{\bngxviii}
\def\LBng{\bngxxii}
\def\hbng{\bngxxv}
\def\Hbng{\bngxxx}
\def\sbns{\bnsviii}
\def\tbns{\bnsvi}
\def\bns{\bnsx}
\def\lbns{\bnsxiv}
\def\Lbns{\bnsxviii}
\def\LBns{\bnsxxii}
\def\hbns{\bnsxxv}
\def\Hbns{\bnsxxx}
\def\sbnw{\bnwviii}
\def\tbnw{\bnwvi}
\def\bnw{\bnwx}
\def\lbnw{\bnwxiv}
\def\Lbnw{\bnwxviii}
\def\LBnw{\bnwxxii}
\def\hbnw{\bnwxxv}
\def\Hbnw{\bnwxxx}



\font\bngv=bang10 scaled 500
\font\bngvi=bang10 scaled 600
\font\bngvii=bang10 scaled 700
\font\bngviii=bang10 scaled 800
\font\bngix=bang10 scaled 900
\font\bngx=bang10
\font\bngxi=bang10 scaled 1100
\font\bngxii=bang10 scaled 1200
\font\bngxiv=bang10 scaled 1400
\font\bngxviii=bang10 scaled 1800
\font\bngxxii=bang10 scaled 2200
\font\bngxxv=bang10 scaled 2500
\font\bngxxx=bang10 scaled 3000

\font\bnsv=bangsl10 scaled 500
\font\bnsvi=bangsl10 scaled 600
\font\bnsvii=bangsl10 scaled 700
\font\bnsviii=bangsl10 scaled 800
\font\bnsix=bangsl10 scaled 900
\font\bnsx=bangsl10
\font\bnsxi=bangsl10 scaled 1100
\font\bnsxii=bangsl10 scaled 1200
\font\bnsxiv=bangsl10 scaled 1400
\font\bnsxviii=bangsl10 scaled 1800
\font\bnsxxii=bangsl10 scaled 2200
\font\bnsxxv=bangsl10 scaled 2500
\font\bnsxxx=bangsl10 scaled 3000

\font\bnwv=bangwd10 scaled 500
\font\bnwvi=bangwd10 scaled 600
\font\bnwvii=bangwd10 scaled 700
\font\bnwviii=bangwd10 scaled 800
\font\bnwix=bangwd10 scaled 900
\font\bnwx=bangwd10
\font\bnwxi=bangwd10 scaled 1100
\font\bnwxii=bangwd10 scaled 1200
\font\bnwxiv=bangwd10 scaled 1400
\font\bnwxviii=bangwd10 scaled 1800
\font\bnwxxii=bangwd10 scaled 2200
\font\bnwxxv=bangwd10 scaled 2500
\font\bnwxxx=bangwd10 scaled 3000


\def\*#1*#2{o\null{#2}{#1}}

\def\d#1{\oalign{\smash{#1}\crcr\hidewidth{$\!$\rm.}\hidewidth}}

\def\sh#1{\setbox0=\hbox{#1}%
     \kern-.02em\copy0\kern-\wd0
     \kern.04em\copy0\kern-\wd0
     \kern-.02em\raise.0433em\box0 }

\begin{abstract}

The Gender Identification (GI) problem is  concerned with determining the gender of the author from a given text. It has numerous applications in different fields like forensics, literature, security, marketing, trade, etc. Due to its importance, researchers have put extensive efforts into identifying gender from the text for different languages. Unfortunately, the same statement is not true for the Bangla language despite its being the 7\textsuperscript{th} most spoken language in the world~\cite{spoken_language}. In this work, we explore Gender Identification from Social media Bangla Text. Specially, we consider two approaches for feature extraction. The first one is Bag-Of-Words(BOW) approach and another one is based on computing features from sentiment and emotions. There is a common stereotype that female authors write in a more emotional way than male authors. One goal of this work is to validate this stereotype for the Bangla language.   
\end{abstract}

\begin{IEEEkeywords}
gender identification, Facebook posts, Bangla language, deep learning, Bag-Of-Words
\end{IEEEkeywords}

\section{Introduction}

Text is the most important means of communication in today's world. Popular online social networking sites such as
Facebook, Twitter, and MySpace etc. are mainly text based. The rapid growth of the Social Media has created enough opportunities to
share information across time and space. Users
 are now comfortable to contribute more to contents of the social media websites and post
their own material. 

The underlying anonymity and the lack of accountability of social media websites make
it difficult to determine the real author of any text posted
online. Any user can
share anything on social media websites and can claim to be another person or
falsely claim to have certain characteristics (such as gender,
age, etc.). To detect this false claim accurately and efficiently, we need an automated tool so that user anonymity is no longer preserved. 

The authorship analysis(AA) problem is concerned with analyzing a text to determine the identity or the characteristics of the author based on content of the text. Determining identity from text is known as authorship authentication problem. Again, determining characteristics of author from text is known as authorship profiling problem. Authorship analysis is achieved based on the context and writing style of the text~\cite{abbasi_chen, altheneyan_bachirMenai,ranjel_rosso,tetreault_cahill}. 

Each type of AA problem has its own applications. One example is from security applications where deception and fraud can be prevented using authorship profiling technique. We can apply authorship profiling techniques to targeted advertisement which is one of the key revenue generator for many Internet based companies. Authorship authentication deals with doubtful attribution to well-known authors such as Shakespeare.

The classical approach to feature extraction for authorship analysis problem is to identify unique stylometric
features of written texts. The underlying assumption here is that each author  has unique writing styles that are relatively fixed and barely changes with time. So we can use stylometric features to uniquely identify the author (or his/her characteristics)~\cite{shalabi_kanaanbt}.\endgraf    

In this work, we are interested in determining gender of author given a text which is basically an authorship profiling
problem. The Gender Identification(GI) problem has numerous applications including from marketing to security. We are interested in addressing the GI problem for social media Bangla text. To the best of our knowledge, only one paper has previously addressed this problem. 

Although gender identification has been widely studied in different languages, it is still under studied in Bangla language. Bangla language is the one of the most widely spoken and
culturally rich language. This language is the 7\textsuperscript{th} most spoken language~\cite{spoken_language} of the world and native language of Bangladesh. However, this is the only reason to study GI problem in Bangla language. The problems associated with the Bangla language and the
relatively under-developed field of Bangla Natural
Language Processing (NLP) makes it more challenging to study
such problems for Bangla.  \endgraf

In this work, we will follow the following steps to study the GI problem. We will use one the dataset collected from one of the previous work. The dataset consists of Facebook text  written by various authors of both genders. We will then pre-process the data and extract feature from the data in 2 ways. One is called Bag-of-words technique and another one is based on computing features related to sentiments and emotions. There is a common stereotype that female authors write in a more emotional way than male authors~\cite{jaffe_lee,Cheng_Na}. One goal of this work is to validate this stereotype. We perform various experiments and do a t-test on the returned result in order to claim or disclaim the statement. \endgraf

The rest of this paper is organized as follows: Section II overviews the related works of gender identification problem. In section III, we have proposed our detailed solution. Section IV presents the experimental results. Finally, in section V, we conclude our findings  with a discussion of the obtained observations and the future directions of this work.

\section{Related Works}
The Gender Identification problem is a binary text classification(TC) problem where the classes are male or female. Thus, we first provide a quick summary of some recent efforts on Bangla TC before we address the associated works. This problem is widely studied in different language. Unfortunately, to the best of our knowledge, gender identification from text is a largely understudied problem in Bangla NLP community. \endgraf

Several papers on Bangla TC have been published in the past decades. Interested reader can check these research papers~\cite{ V_K_Singh, Islam_hossain,paul_shill}. Majority of the current works in these papers collected datasets collected various websites including social media websites, news websites etc. and follow the general approach
as follows. They pre-process the data to remove unwanted contents like punctuation, stop words, numbers and non-Bangla words. Then feature vector is computed using different weight scheme of Bag-Of-Words approach or any other approach such as n-gram words/characters. Sometimes Part-of-speech(POS) tagging, stemming and other feature reduction techniques are applied to minimize the large number of computed features. Finally, several classification algorithms are trained on some of the dataset and tested on the rest of the dataset. 

The Bangla TC problem has recently been addressed in a number of intriguing research that are important to authorship analysis problems. One example is in ~\cite{tripto_ali} where authors focus on the new writing styles of YouTube comments. They use Word2vec algorithm and TF-IDF as feature extraction techniques and apply deep learning method and traditional machine learning model to identify sentiments from Youtube comments. Finally, they compare deep learning model performance against traditional machine learning performance. 

Compared to TC, AA problem is still largely understudied problem in literature for Bangla language to the best of our knowledge. Below we discuss some works of authorship attribution before going to more relevant works of gender identification problem. For both problems, computing stylometric features is common as It is thought to be a more intuitive strategy than the BOW approach. 

To the best of our knowledge, the authorship attribution problem is largely understudied problem for Bangla language as the number of published work is very small~\cite{das_mitra}. Das and Mitra performed authorship attribution in 36 documents and 3 authors by extracting uni-gram and bi-gram features. Chakraborty~\cite{tanmoy_chakraborty} worked with stylometry feature on top of SVM model for 3 authors and achieved 84\% accuracy. Shanta Phani et. al. ~\cite{Phani_Lahiri} used six types of lexical n-gram features on 3 authors with machine learning model. Das et al. ~\cite{das_tasnim} extracted hand-drawn features
such as word frequency, type-token ratio, number of various
POS, word/sentence lengths etc. and applied these features on 4 authors. Hossain et. al. 
~\cite{Hossain_Ismail_Islam} scrutinized six columist of current times and applied stylometry feature to observe writing patterns of these columnists. They got 90.67\% accuracy rate on three hundred test articles. Phani,
Lahiri, and Biswas  ~\cite{phani_lahiri} conducts experiments by employing  character bi-gram and tri-gram as feature extraction and got 99\% accuracy using Multi layered Perceptrons model. Chowdhury et. al. employed work embeddings and concluded that fastTexts algorithm with CNN outperforms all other models. Khatun et. al. ~\cite{Khatun_rahman} employed character embedding on 2 dataset from 6 to 14 authors and got the improved accuracy by 10\%.    

In authorship profiling problem, we are interested to identify one or more characteristics of author. Specifically, in gender identification problem, we are interested to identify gender of author from text. Researchers in ~\cite{peng_circone} have used a set of most frequent n-gram features to identify an author's gender from text. Researchers~\cite{moshe_shlomo} used  Multi-Class Real Window (MCRW) learning algorithm to identify authors gender from blogs. They obtained 86\% average accuracy    using a corpus with a size of 71,000. Authors in ~\cite{lim_goh} applied 
Support Vector Machine classifier together with Principal Component Analysis for age and gender profiling in 2013. Style-based and content-based features are used in this research  with average accuracy 82.6\%.
Shoev et. al.~\cite{sboev_rybka} used a set of morphological and syntactic features to identify gender and sentiment from Russian texts. They used LSTM and GRU deep learning models. They have applied Gradient Boosting, adaBoosting, ExtraTrees, Random Forest, PNN (sigma=0.1), SVM, ReLU (1 Hidden Layer), CNN+MLP, CNN+LSTM machine learning algorithm. The best achieved accuracy is \SI{86 \pm 3}{\percent} with combination of CNN and LSTM algorithm. 
Alsmearat et. al. ~\cite{shalabi_kanaanbt} used articles from online newspaper websites to extract author gender. They have employed bag-of-words and stylometric features. By using decision tree algorithm They achieved best average 77.8\% accuracy. 
Zrigui. et. al.~\cite{Bsir_zrigui} used Gated Recurrent Unit Deep learning model to detect gender from Facebook and PAN corpus. For feature extraction, they used lexical features, n-grams, Bag of smileys, stop words, etc. Their achived  accuracy for Facebook and PAN corpus are 62.1\% and 79\% respectively.  Cheng et. al. ~\cite{Cheng_Na} used stylometric features to  identify genders from English data set including email of institutions and the online newspaper. They used traditional machine learning algorithms such as SVM, Adaboost Decision Tree, and logistic regression. The  best achieved average accuracy is from SVM model for this problem which is 83\% respectively. Researches in ~\cite{rahman_kabir}applied  stylometric features to identify genders from Bangla speech data. from a medium-scale Bengali speech corpus of the Bengali Daily Newspaper Prothom Alo. Traditional machine learning algorithm is used for this problem. They have got average 99.9\% precision, 100\% recall and 99.9\%  F1 score. \endgraf 

 Finally, researchers~\cite{Alsmearat_Kanaan_Shalabi} haveidentified authors gender from newspaper articles. They have also found impact of emotion in identifying genders. They have used Bag-Of-Words and Sentiment/Emotion lexicons as features in traditional machine learning algorithm like SVM, NB, Decision Tree \& KNN models. The best achieved accuracy 86.4\% for SVM with Bag-Of-Words feature.

\section{Methodology}
\begin{figure}[htp]
    \centering
    \includegraphics[width=\columnwidth]{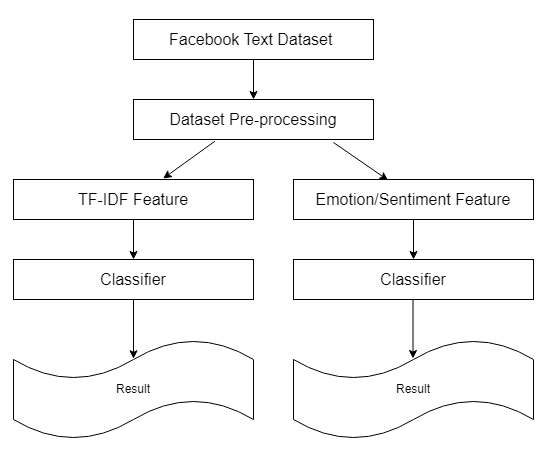}
    \caption{Steps followed in our solution}
    \label{fig:3.methodology}
\end{figure}

This section presents a detail overview of two feature extraction techniques. One is Bag-Of-Words feature and another one is Sentiment/Emotion feature approach. We first label the dataset and then compute the feature after pre-processing the data. The feature is then feed into machine learning model. Then we provide the architecture of the model.    Fig~\ref{fig:3.methodology} presents an overview of our proposed solution.  

\subsection{Dataset Creation}\label{AA}

Like any other TC problem, the first step in Gender Identification problem is to build a large dataset. 
We have collected the dataset from Github~\cite{sultan2023}.This dataset contains Facebook posts with associated gender information. Dataset in this repository is created from 3 Facebook groups which are listed in Table~\ref{facebook_group_name_table}.\endgraf

\begin{table}[htbp]  
\caption{Facebook Group for Data Collection}
\begin{center}
\begin{tabular}{|p{0.2\textwidth}|p{0.2\textwidth}| }
\hline
 Group Name & Group Link
 \\ 
 \hline
 
1.  {\bng BueyeT Aairhepet eshana } 
& 
\href{https://www.facebook.com/groups/OverheardAtBUET}{OverheardAtBUET}
 \\ 
 \hline
2. BCS Preparation for BUET Students &
\href{https://www.facebook.com/groups/BCS.Aspirants.BUETian}{BCS.Aspirants.BUETian}
 \\ 
 \hline
3. {\bng EI papii ecaekh BueyeT Ja edekhich } 
&
\href{https://www.facebook.com/groups/895957014215386}{link}
 \\  
 \hline
\end{tabular}
\label{facebook_group_name_table}
\end{center}
\end{table}

The author~\cite{sultan2023} collects Facebook posts with  author names. All the facebook group is related to Bangladesh University of Engineering and Technology where there are less female students (20\%-30\%) than the male students. That is why, the number of female data is low in the dataset. The dataset contains 3896 posts in total out of which 2947 are from male authors and 949 are from female authors. 
 

\subsection{Pre-processing}
The post obtained from Facebook is noisy and often contains a lot of unnecessary information. Here URL, image, tags etc. are present in the text. So we apply various pre-processing steps and have filtered out all characters except Bangla characters. Then we tokenize our texts and remove stop words from the text. We collect Bangla stop words from Github repository~\cite{git_stop_word_bangla} as mentioned in research work~\cite{tripto_ali}. 

Elongated words often contains some context in identifying gender data from text. We express our feelings in elongated words. For example, "Greaaat news!!!" has more feelings than "Great news!!!". Generally, female tends to express their feelings by elongated words. To maintain the context of the text, we do not apply lemmatization. 
\subsection{Word Vector Formation}
We use traditional machine learning models in our proposed solution. To use these models, we need to convert our text to a word vector. Conversion to word vector from text is done using TF-IDF features and Sentiment/Emotion Features approaches.    
\begin{enumerate}[wide, labelwidth=!, labelindent=0pt]

\item   
\newcolumntype{s}{X}
\newcolumntype{b}{>{\hsize=1.5\hsize}X}

\begin{table*}[t]
 \caption{Example Sentence for TF-IDF Vectorizer 
}
\label{tf-idf-text-example}
\begin{tabularx}{\textwidth}{sb}
\toprule
Text No. & Example\\
\midrule
Text1 &  I love natural language processing.\\ 
Text2 & I like data processing. \\ 
Text3  & I like text processing. \\ 
\bottomrule
\end{tabularx}
\end{table*}

\begin{table*}[t]
 \caption{Example Sentence for TF Vectorizer 
}
\label{tf-vectorizer-example}
\begin{tabularx}{\textwidth}{|s|s|s|s|s|s|s|s|s|}
\toprule
Text No. & I & love & like & natural & language & data  & text  & processing\\
\midrule
Text1 &  1 & 1 & 0 & 1 & 1 & 0 & 0 & 1 \\ 
\midrule
Text2 & 1 & 0 & 1 & 0 & 0 & 1 & 0 & 1  \\
\midrule
Text3  & 1 & 0 & 1 & 0 & 0 & 0 & 1 & 1 \\ 
\bottomrule
\end{tabularx}
\end{table*}

\begin{table*}[t]
 \caption{Example Sentence for IDF Vectorizer 
}
\label{idf-vectorizer-example}
\begin{tabularx}{\textwidth}{|s|s|s|s|s|s|s|s|s|}
\toprule
No. & I & love & like & natural & language & data & text & processing\\
\midrule
Word &  0 & 0.477 & 0.18 & 0.477 & 0.477 & 0.477 & 0.477 & 0 \\ 
\bottomrule
\end{tabularx}
\end{table*}

\begin{table*}[t]
 \caption{Example Vector for TF-IDF Vectorizer 
}
\label{tf-idf-vectorizer-example}
\begin{tabularx}{\textwidth}{|s|s|s|s|s|s|s|s|s|}
\toprule
Text No. & I & love & like & natural & language & data & text & processing\\
\midrule
Text1 &  0 & 0.477 & 0 & 0.477 & 0.477 & 0 & 0 & 0  \\ 
\midrule
Text2 & 0 & 0 & 0.18 & 0 & 0 & 0.477 & 0 & 0  \\
\midrule
Text3  & 0 & 0 & 0.18 & 0 & 0 & 0 & 0.477 & 0 \\ 
\bottomrule
\end{tabularx}
\end{table*}

TF-IDF Count Vectorizer Approach: \endgraf

\hspace{10pt} TF-IDF stands for Term Frequency Inverse Document Frequency. TF-IDF converts the text of user into a meaningful numbers of vector to fit in the machine learning model. In any document, the term frequency is the number of occurrences of a term . On the other hand, document frequency represents the number of documents containing of that term.  Term frequency indicates the importance of a specific term in a document. Document frequency indicates how common the term is in~\cite{tf_idf_link}. TF-IDF gives higher weights to terms appearing frequently
in the given document and rarely in other document.

We have implemented TF-IDF count vectorizer from scikit-learn library. Since, TF-IDF generates higher dimensional feature vectors, we need dimensional reduction techniques. We can do these in several ways. First one is to reduce the number of extracted terms. Another one is to use stemming. In our work, we have used the reduction of extracted terms approach. We set the number of extracted terms as 1000.    

For Example, let us take an example. Suppose, we have 3 texts which have to be converted to vector using TF-IDF count vectorizer as sho0wn in the table Table~\ref{tf-idf-text-example}. 

For converting these texts into vector, we have to first identify unique words and count how many times these words occur in each text. This is shown in Table~\ref{tf-vectorizer-example}

Then we have to compute inverse document frequency (IDF) using the following formula. 

$$ idf_i = log\frac{n}{df_i} $$

where $df_i$ represents how many documents contain term i and n is the total number of documents. We have calculated the inverse document frequency for each work and shown it to Table~\ref{idf-vectorizer-example}

Now we will multiply TF matrix with IDF score to get vectorized form of each text which is shown in Table~\ref{tf-idf-vectorizer-example}. We have converted all texts into vector. These vector can be feed in any machine learning algorithm.

\item   
Sentiment/Emotion Feature Extraction Approach: \endgraf

\hspace{10pt} For Sentiment/Emotion feature, we explore NRC Emotion Lexicons (EmoLex) provided by Mohammad and Turney~\cite{Mohammad_turney}. There are ten unweighted lexicons in EmoLex. Two of them are for sentiments(positive, negative) and rest of them are containing emotions which are anger, anticipation, disgust, fear, joy, sadness, surprise and
trust. 

EmoLex conatins sentiment and emotional words for several languages like English, Bangla, Arabic etc. We collect the Bangla sentiment and emotion words from the list.
Then We compute
a feature for each sentiment/emotion in each Facebook Text. To do this, we take the sum of the numbers of occurrences
of all terms the lexicon associated with the sentiment/emotion
under consideration.

\end{enumerate}
\subsection{Model Architecture}
For two types of features, we have implemented deep learning model and traditional machine learning model to extract gender from text. In deep learning model, we implement LSTM and GRU model. In traditional machine learning model, we implement SVM, NB, Decision Tree and KNN algorithm. 

\begin{figure*}[ht!]
       \centering
\begin{subfigure}[t]{0.8\textwidth}
\includegraphics[width=\textwidth]{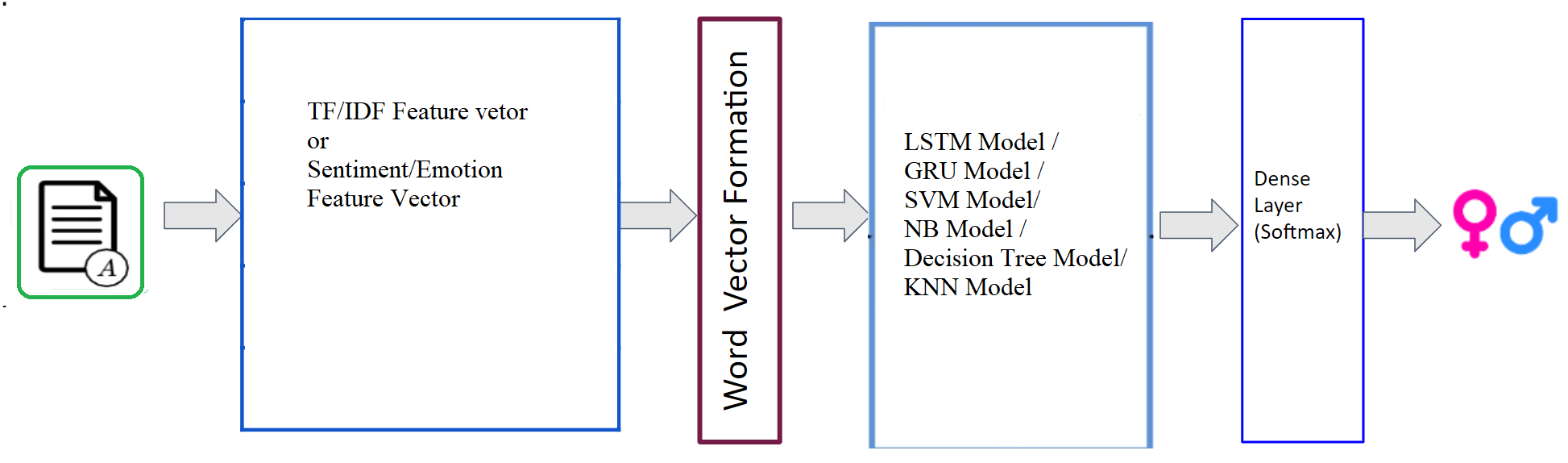}
 
\end{subfigure}
\bigskip
\caption{Architecture for Gender Identification from text}
  \label{fig:model_architecture}
\end{figure*}

 Fig.~\ref{fig:model_architecture} shows the architecture of LSTM model. Traditional models such as Support Vector Machine and Naive Bayes are implemented to evaluate the performance of the deep learning model. 

\begin{enumerate}
  \item LSTM and GRU model with TF-IDF Features: \\
     We have prepared word vectors by TF-IDF features discussed in Section III-C. We pass these vectors to the LSTM layer having 300 node. The output of the LSTM layer is passed to a dense layer. Softmax~\cite{sharma_ochin} is used as an activation function. The optimizer is RMSprop and binary cross entropy ~\cite{shipra_saxena} is used as a loss function. The same process is repeated for Gated Recurrent Unit(GRU) model. We have noted down each model accuracy, Precision, Recall, F1-score and AUC.
     Fig. \ref{fig:model_architecture} shows the architecture of LSTM and GRU model with the stylometric feature.    
  
  \item LSTM and GRU model with Sentiment/Emotion Feature Vector:\\
For any sentence s with classification c, we have calculated the sentiment/emotion feature vector.  This vector is then passed through a LSTM layer. The output of LSTM layer is passed to a dense layer which is used to detect gender. The same process is repeated for Gated Recurrent Unit(GRU) model. For each model,we have noted down each model accuracy, precision, recall, F1-score and AUC.  Softmax ~\cite{sharma_ochin} is used in dense layer as activation function because each sentence can belong to only one of two classes in our scenario. The optimizer is RMSprop and binary cross entropy ~\cite{shipra_saxena} is used as a loss function.
    
  \item Traditional Models:\\ 
  We have implemented traditional machine learning algorithms such as Support Vector Machine(SVM) and Naive Bayes (NB), Decision Tree and KNN to identify gender from the text. We have  used Sentiemnt/Emotion lexicons and Term Frequency Inverse Document Frequency (TF-IDF) to generate feature-set from the text. 2 gram is used in the TF-IDF vectorizer. We have used linear kernel function in the Support Vector Machine model. The function of kernel is to take data as input and transform it into the required form. For each model, we have noted down accuracy, precision, recall, F1-score and AUC. 
\end{enumerate}

\section{Experimental Evaluation}
In this section we have done a  set of experiments in order to evaluate
the efficiency of
different classifiers on the collected dataset. We have also evaluated efficiency of different feature selection techniques. 

These experiments are conducted on Python integrated environment. We have used Tensorflow framework for deep learning model implementation. Scikit-learn is used to implement traditional machine learning model. The experiment is conducted on a machine with Intel Core i7 processor with 1.8 GHz clock speed and 16 GB RAM. The machine has also an NVidia Geforce 150 with 2 GB memory and therefore Tensorflow based experiment can utilize GPU instructions. 

For the testing option, we first split the dataset into 10 groups. For each group, we have taken 80\% as training data, 10\% as testing data and 10\% as verification data. For each group, the accuracy, precision, recall (sensitivity), F1-score
and Area Under the ROC Curve (AUC) is measured. Then these accuracy measures are averaged up. 

We conduct the experiments in three ways. Firstly, only the feature with TF-IDF vectorizer is involved in the experiment while, in
the second way, the experiment is conducted with the SE features alone.
Finally, experimentation is conducted by combining both feature sets. Table~\ref{tf_idf_feature_along} shows the results for the BOW features
alone. The table shows the superiority of the SVM overall classifiers. This is expected as Bayesian classifiers and SVM perform very well for the general TC problem~\cite{Alwajeeh_Ayyoub_Hmeidi,Marton_Wu}. The table also shows that LSTM, GRU, DT,  NB, and KNN perform well.      

\begin{table}[htbp]  
\caption{RESULTS FOR THE TF-IDF FEATURES ALONE}
\begin{center}
\begin{tabular}{|p{0.058\textwidth}|p{0.058\textwidth}| p{0.058\textwidth}|
p{0.058\textwidth}|
p{0.058\textwidth}|
p{0.058\textwidth}|}
\hline
  & Acc & Pre & Rec & F1 & AUC
 \\ 
 \hline
LSTM & 76.41\%  & 78.15\% & 95.22\% & 85.85\% & 57.4\%
 \\ 
 \hline
GRU & 76.02\% & 78.03\% &  94.31\% & 85.4\% & 58.65\%
 \\ 
 \hline
SVM & 77.56\%  & 78.21\% & 97.64\% & 86.85\% &  56.00\%
 \\ 
 \hline
NB &  76.03\% &  76.09\% &  98.97\%  & 86.03\% & 55.89\% 
 \\ 
 \hline
DT & 72.95\% &  79.45\% & 87.06\% &  83.08\%  & 57.31\% 
 \\ 
 \hline
KNN &  68.21\%  & 77.10\% & 81.85\% & 79.40\% &  54.70\% 
 \\ 
 \hline
\end{tabular}
\label{tf_idf_feature_along}
\end{center}
\end{table}

Table~\ref{emotional_sentiment_along} shows the results for the Sentiment and Emotion features alone. The table shows a surprising trend that relying on such features
produces results much better than the random baseline in the case of some models.
This means that these features possess some importance
suggesting that there might be little or some difference between
male and female authors in terms of the use of sentiment baring
or emotional terms.  

\begin{table}[htbp]  
\caption{RESULTS FOR THE TF-IDF FEATURES ALONE}
\begin{center}
\begin{tabular}{|p{0.058\textwidth}|p{0.058\textwidth}| p{0.058\textwidth}|
p{0.058\textwidth}|
p{0.058\textwidth}|
p{0.058\textwidth}|}
\hline
  & Acc & Pre & Rec & F1 & AUC
 \\ 
 \hline
LSTM & 72.77\%  & 75.37\% & 92.84\% & 83.20\% & 56.17\%
 \\ 
 \hline
GRU & 78.93\% & 78.10\% &  92.80\% & 84.82\% & 59.10\%
 \\ 
 \hline
SVM & 72.64\%  & 76.82\% & 90.64\% & 83.16\% &  55.27\%
 \\ 
 \hline
NB &  76.88\% &  76.88\% &  100.0\%  & 86.92\% & 50.0\% 
 \\ 
 \hline
DT & 64.02\% &  75.64\% & 75.78\% &  75.71\%  & 53.15\% 
 \\ 
 \hline
KNN &  70.18\%  & 76.37\% & 85.79\% & 80.81\% &  56.67\% 
 \\ 
 \hline
\end{tabular}
\label{emotional_sentiment_along}
\end{center}
\end{table}

\section{Conclusion \& Future Work}

In this work, we have automatically identified the gender of the author from Social media Bangla text. This is an important but largely understudied problem that has numerous applications including security such as fraud detection in social media to marketing. We explored the Gender Identification problem for Bangla text as a supervised learning problem. Specifically, we have compared two feature-based approaches for deep learning and traditional machine learning algorithms. The first one is the TF-IDF approach
while the second one is based on computing features based on
sentiments and emotions. One goal of this work was to confirm
the validity of the common stereotype that female authors tend
to write in a more emotional way than male authors. Our
results showed that this is somewhat true for the Bangla dataset.


\printbibliography

\end{document}